\documentclass[twocolumn,trackchanges]{aastex701}
\usepackage{placeins}
\usepackage{amsmath}

\begin{document}

\title{Super-Chandrasekhar white dwarfs by the evolution of magnetized main-sequence stars: New mass-limits from STARS simulation}

\author[0009-0000-6980-6334]{Zenia Zuraiq}
\email[show]{zeniazuraiq@iisc.ac.in}
\affiliation{Department of Physics, Indian Institute of Science, Bengaluru 560012, India }

\author[0000-0002-3020-9513]{Banibrata Mukhopadhyay}
\email[show]{bm@iisc.ac.in}
\affiliation{Department of Physics, Indian Institute of Science, Bengaluru 560012, India }

\author{Achal Kumar}
\email{achal.kumar@ufl.edu}
\affiliation{Department of Physics, University of Florida,
Gainesville, Florida, USA}

\author[]{Arnab Sarkar}
\email{arnabsarkar1510@gmail.com}
\affiliation{Institute of Astronomy, University of Cambridge, UK}

 \author[0000-0002-8277-2033]{Alexander J. Hackett}
\email{ajh291@cam.ac.uk}
\affiliation{CEICO, Institute of Physics of the Czech Academy of Sciences, Czechia}

 \author[0000-0002-6389-2697]{Projjwal Banerjee}
\email{projjwal@iitpkd.ac.in}
\affiliation{Indian Institute of Technology Palakkad, Kerala, India}

\author[0000-0002-1556-9449]{Christopher A. Tout}
\email[show]{cat@ast.cam.ac.uk}
\affiliation{Institute of Astronomy, University of Cambridge, UK}

\begin{abstract}
We explore the time evolution of the magnetized main-sequence stars (MSs) to
magnetized non-rotating white dwarfs (WDs) with a series of models, capturing modified Hertzsprung-Russell diagram. We use the
Cambridge Stellar evolution code, STARS, with  modifications by introducing magnetic effects and cooling.  We investigate the possible
existence of stable, magnetized, massive, super-Chandrasekhar, approximately spherical WDs,
away from Chandrasekhar's mass--radius curve.  These we refer to as
B-WDs.  They are inferred from the observations of peculiar
over-luminous type~Ia supernovae, e.g. SN2003fg.
We show that the classical effects of
magnetic fields significantly impact MSs and their time evolution to WDs. 
Moreover, the magnetic field could be decisive in explaining the unusually large radius observed in some sub-Chandrasekhar WDs, e.g. SDSS J123204.19+522548.2, along with the finite temperature effect.
We further investigate the stability
of magnetized WDs, both below and above the Chandrasekhar limit and
confirm the possibility of super-Chandrasekhar B-WDs with new mass
limits that depend on the magnetic field geometry, mass accretion from
a companion and cooling rates. Therefore, while the “Chandrasekhar limit” may be sacrosanct, its value severely depends on the underlying physics, here the magnetic field.

\end{abstract}


\section{Introduction} 
\label{sec:intro}

One of the most celebrated discoveries of 20th century astrophysics
is the Chandrasekhar mass-limit ($M_{\rm Ch}$) for white dwarfs \citep[WDs:][]{chandralimit}.  The standard $M_{\rm Ch} = 1.44M_\odot$ applies to
slowly rotating, weakly magnetized WDs. This relates
to type~Ia supernova(e) (SNIa/SNeIa), which are important standard candles used
to understand the expansion of the Universe,
directly influencing our understanding of modern cosmology.

As a cold carbon--oxygen (CO) WD approaches its limiting mass, carbon can
ignite degenerately at its core.  Without other influence, this occurs
at about $1.38M_\odot$, just shy of $M_{\rm Ch}=1.44M_\odot$, when the core density ignites carbon fusion.  The
ensuing thermonuclear runaway converts $\sim1M_\odot$ of the WD
material to iron group elements, typically nickel-56 ($\!^{56}$Ni), in nuclear
statistical equilibrium.  The nuclear energy released explodes the WD
and a SNIa is powered by the radioactive decay of
$\!^{56}$Ni to iron-56 ($\!^{56}$Fe).  
The fixed ignition mass makes
SNeIa calibrateable standard candles and, thence, an important rung on
the cosmic distance ladder. Indeed, most SNeIa obey a fixed
luminosity--light-curve stretch (Phillips) relation \citep{philips}.  However,
there also exist classes of peculiar under- and over-luminous SNeIa,
which do not follow the Phillips relation \citep{Taubenberger2017}.
While some under-luminous SNeIa predict progenitor masses as low as
$0.5M_\odot$ \citep{underlum}, some over-luminous SNeIa predict
progenitor masses as high as $2.8M_\odot$
\citep[e.g. SN2003fg, SN 2009dc:][]{sn2009dc,SN2003fg}.  Questions are: (1)~are all SNeIa safely
standardizable candles; and (2)~is $M_{\rm Ch}$
sacrosanct?  While we cannot deny observations, how robust are the
measurements (which are also model-based) and their predictions? (3) Finally, what is the source of significant violation of $M_{\rm Ch}$?

WDs have long been understood to violate Chandrasekhar limit leading to super-Chandrasekhar mass (super-CM) in the presence of magnetic field, however, without considering their stability \citep{ostriker,adam}. Studies of stable, magnetized WDs with significantly super-CM and sub-CM limits and their astrophysical implications have been initiated by our group \citep{udas,sathya,surajit2018,deb}.

Although the effect of rotation can also contribute to super-CM WDs, it may still be sub-dominant when compared to the magnetic field effect, depending on the chosen field strength \citep{sathya,franzon_rotwd,komatsu}. Merger of two sub-CM WDs could, in principle, lead to super-CM WD, however they may face off-center carbon burning or collapse to a neutron star in place of producing stable, massive progenitor and eventually a SNIa \citep{saio_offcenterC,nomoto_ns,timmes,hachisu}. However, other effects, including modified gravity, violating $M_{\rm Ch}$ were demonstrated \citep{udas_modified,ungravity,charged,fRT_gravity}.  

Magnetic or highly magnetic WDs have many other observable properties
depending on the misalignment between their rotation and magnetic
axes, i.e. obliquity angle. They may behave as pulsars \citep{wdp1,wdp2,wdp3} and
also be a potentially important source for the LISA and futuristic
IndIGO-D/DECIGO/BBO gravitational wave (GW) detectors \citep{mayusree_wd, surajit2019}. Therefore, we may have direct
detection of magnetized WDs which deviate from the Chandrasekhar
mass-radius ($M-R$) relation and reach super-CMs.  We
call these B-WDs.

Previous models of B-WDs are based on stand-alone solutions of
Einstein--Maxwell equations.  In general they are solutions to the
Tolman--Oppenheimer--Volkoff equation with magnetic fields and
rotation added \citep{sathya,surajit2019,franzon_rotwd}.  

The following are important questions.
\begin{enumerate}
\item Can magnetized main-sequence 
stars (MSs) lead to magnetized
  super-CM WDs after giant evolution and WD cooling?
\item What then would be the tracks followed in an Hertzsprung--Russell
  (H--R) diagram?
\item Is stability of such a magnetized, massive star guaranteed?
\item Is the magnetic field required to significantly violate the
  $M_{\rm Ch}$ achievable when the star arrives at its WD
  phase?
\end{enumerate}

In order to answer these questions, we explore detailed models of the time evolution
of magnetized MSs to magnetized WDs with the Cambridge
stellar evolution code STARS \citep{eggleton,eggleton2,pols}. Following previous work predicting the end phase of a MS to be toroidally dominated \citep{tout_mostmagnstars}, we consider toroidally dominated stars, which hardly deviate from spherical symmetry if the inner magnetic field is restricted to $\sim 10^{14}$ G (see, e.g., \citealt{sathya}). This is in accordance even with stability criteria based on barotropic assumption suggesting that the toroidal field could be 30 times the poloidal field \citep{BR2009,mayusree2}. This also confirms that stronger fields, which could deviate the star from spherical symmetry, are unphysical. Hence,
one-dimensional STARS code suffices for the present purpose. There are two 
pathways for MSs  to evolve to WDs that
are magnetized and super-CM. 
\begin{itemize}
\item \textit{Single-star evolution}: A sufficiently massive,
  magnetized MS evolves to a WD which, with a sufficiently strong
  magnetic field, could be super-CM, depending on the
  properties of the MS.
\item \textit{Binary-star evolution with mass transfer}: A weakly
  magnetized WD, typically sub-CM, forms from the evolution
  of a magnetized MS and later accretes matter from a binary companion.
  The process of accretion strengthens the magnetic field allowing
  the degenerate matter to support more mass leading to a super-CM WD.
\end{itemize}
Here we focus on the second scenario.  
There are many routes to reach the end points of accreting 
WDs that we model here. Typically, a more massive MS evolves to a
giant or asymptotic giant branch (AGB) star, filling its Roche lobe on the way, and transfers much of its envelope to a less massive companion, exposing its degenerate core. This core cools to the WD, while the other one now becomes the more massive companion, which evolves and transfers its envelope back to the WD. However, the exact route is not our focus here, because we know that it is possible to reach the SNIa ignition mass by accretion, so with magnetic support the mass can be exceeded.
Also our exploration of detailed single-star models will be reported elsewhere.

Our models probe the possible future trajectories of currently
observed magnetic WDs.  We find that, through accretion on to a present-day magnetized WD, a super-CM 
B-WD could form in future.

This is the first approach of its kind to explore the time evolution of
magnetized MSs into B-WDs. Our approach, as we demonstrate below, leads to significant results on a few levels by (i) establishing a new magnetized stellar evolution pathway to the formation of massive WDs, (ii) showing a modified H–R diagram in presence of magnetic field and accretion. We illustrate how the WD $M-R$
relation deviates over time from that of Chandrasekhar and approaches
a new super-CM branch.  We confirm that B-WDs may exist
in reality, and have direct effects on the $M-R$ relation even at the sub-CM level \citep{drisya,parsons}, motivating observers to search directly for them.

\section{A formation scenario for super-Chandrasekhar B-WDs.} 
\label{sec:form}

Using STARS, we first evolve magnetized MSs to magnetized (CO and helium-CO: HeCO) WDs with dormant magnetic fields through the AGB
phase.  We describe this process in Appendix B.

By dormant magnetic field, we mean a dynamically unimportant magnetic field, i.e. the central field does not alter the evolutionary track
leading to WD formation and does not cause it to deviate from the
Chandrasekhar $M-R$ relation. This remains the case until the field $\lesssim5 \times 10^{13}\ G$.

\begin{figure}[!htpb]
	{
	\includegraphics[scale=0.35]{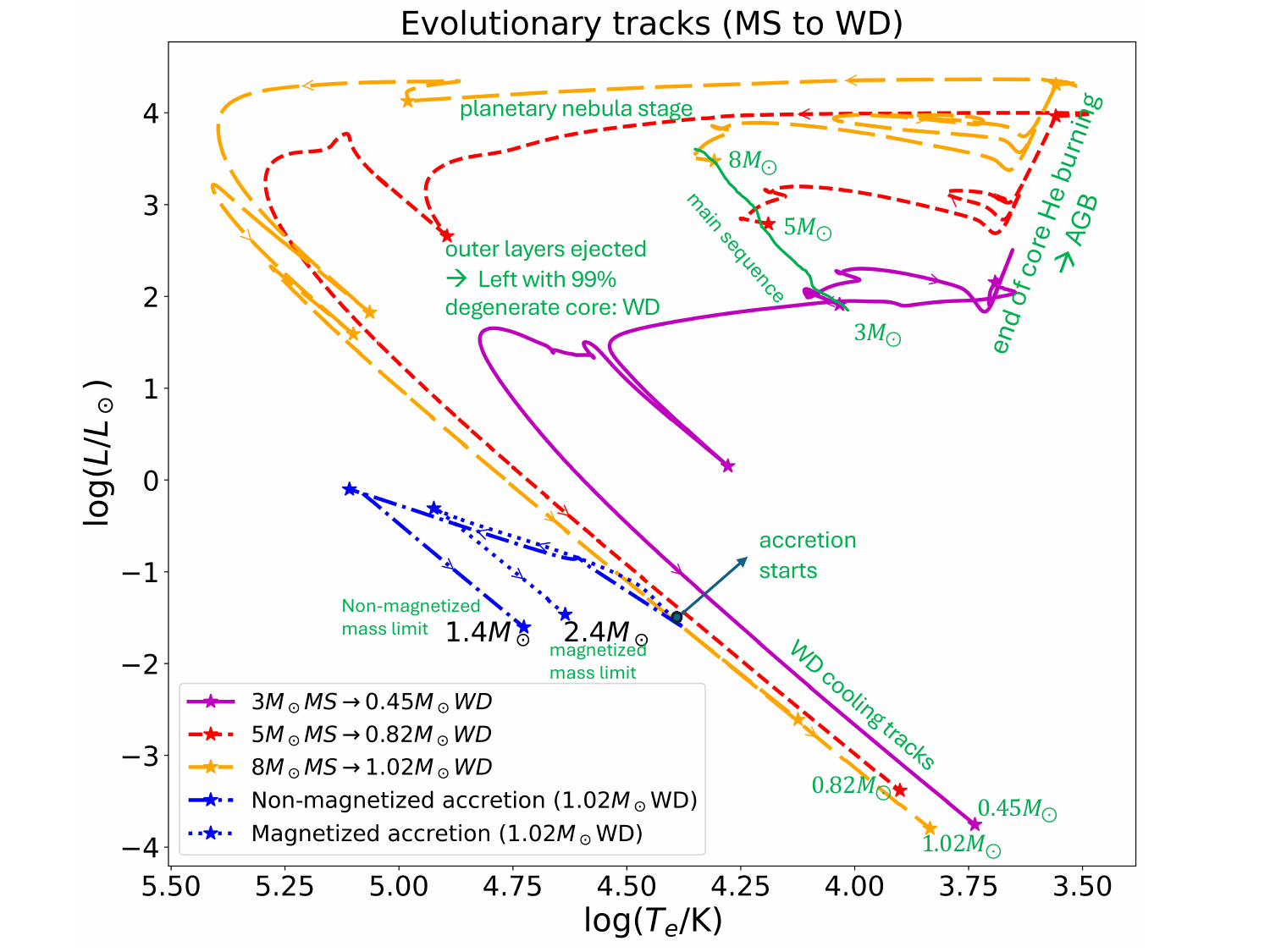}}
	\caption{Evolution of MSs of various masses to WDs in
          H--R diagrams, where $L$ is the luminosity and $T_{\rm e}$
          the effective temperature of the star. The solid, dashed and long-dashed lines show the evolution from MS to WD without
          accretion, from bottom to top: for $3,~5$ and $8M_\odot$ MSs.   
          Accretion at $\dot{M} = 10^{-9}\,M_\odot\,{\rm yr}^{-1}$ on to the $1.02M_\odot$ WD
          with an initially dormant magnetic field leads to a $2.4\,M_\odot$ B-WD (dotted line), whereas without magnetic field forms a $M_{\rm Ch}=1.4M_\odot$ WD (dash-dotted line). The magnetic evolution is done as per the route described in the Appendix with magnetic field parameters $\eta = 1.3, \gamma = 0.9$.}
    
    \label{hr}
\end{figure}

Fig.~\ref{hr} shows typical examples of the underlying H--R diagram. 
We further explore the
super-CM B-WDs originating from dormantly magnetized WDs
undergoing accretion and how it differs in the absence of a magnetic field.

We model spherically symmetric accretion on to the dormantly magnetized WD with STARS. As the central density goes above a critical density due to accretion, the so far dormant field becomes strong enough for the WD to support a significantly larger total mass at a
given radius.  Eventually, the mass can exceed $M_{\rm Ch}$. 
In the absence of
magnetic fields (or below the threshold field), the non-magnetized (or
dormantly magnetized) WD would have evolved up to  $1.38M_\odot$, close to the $1.44M_\odot$ $M_{\rm Ch}$ for an ideal case, beyond which carbon ignition is triggered in the core. Following the accretion up to saturation masses, the WD/B-WD continues on its cooling track. Fig.~\ref{hr} shows how accretion on a fairly hot
$1.02M_\odot$ CO WD \citep[evolved from a $8M_\odot$ MS:][]{IfmR_Binaries} changes the evolution compared to the case with no accretion. In the presence of initial dormant fields, the accretion of rate $\dot{M}=10^{-9}\,M_\odot\,{\rm yr}^{-1}$ leads to a $2.4M_\odot$ super-CM limit (when the dormant field becomes strong). 
See Appendix for the details of the chosen magnetic field, whose profile, in this case, corresponds to $\eta =1.3$ and $\gamma = 0.9$.

\begin{figure}[!htb]
\includegraphics[width=0.45\textwidth]{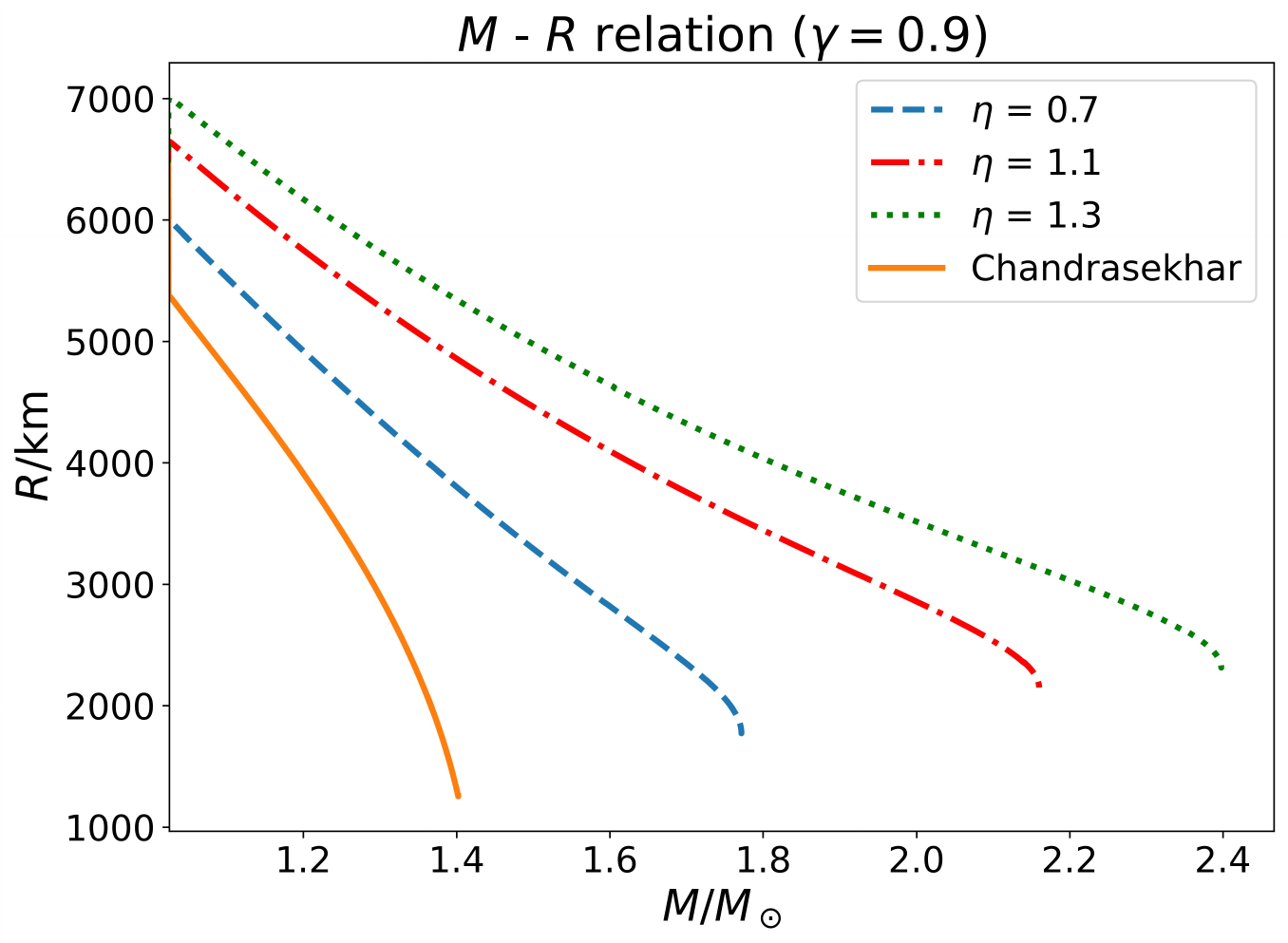}
\caption{$M-R$ curves showing different mass
          limits for different magnetic field profiles, i.e. different
          $\eta$ ($\eta$ and $\gamma$ describe the magnetic field profile defined in the Appendix A). $\dot{M} =
           10^{-9}\,M_\odot\,{\rm yr}^{-1}$ began at $1.02\,M_\odot$ for
          all cases. }
    \label{MR}
\end{figure}

We thus obtain a modified $M-R$ relation and altered maximum WD mass
caused by the effects of the magnetic field.  Fig.~\ref{MR} shows a
series of $M-R$ relations for different magnetic field profiles, 
interestingly with a series of new mass limits, in contrast to Chandrasekhar's case.

We use a density-dependent magnetic field profile throughout the star leading to approximate flux freezing\footnote{See Appendix C showing that the temperature remains almost constant, hence reflecting high conductivity, in almost the entire WD.}, i.e. the magnetic field gets stronger during the stellar contraction leading to the denser parts of the evolution. The central magnetic field typically rises from $\sim 10^{12}\,$G in the
initial low-mass WD to $\sim 10^{14}\,$G.  It reaches $8\times 10^{13}\,$G
for the final $2.4\,M_\odot$ WD in Fig.~\ref{hr}. Throughout, the magnetic to gravitational energy ratio of B-WDs and MSs is maintained \citep{UniversePaper} $\lesssim 10^{-1}$ and $\lesssim 10^{-5}$, respectively, confirming stability \citep{BR2009}, while indeed our time evolution continues uninterrupted unless mentioned explicitly.

The surface field $B_{\rm s}$ is determined by the surface
magnetic flux of the initial MS through explicit flux freezing ($B_sR^2 = $ constant). This is not a bad assumption given that temperature remains almost constant, particularly for smaller B-WDs, except very close to the surface. See Appendix and the figure therein for more details of the magnetic field profile.

Depending on the magnetization of the initial MS, we work
with initial WDs with $B_{\rm s}$ of about $10^4\,$G to $10^7\,$G, consistent with observations \citep{ferrario}, corresponding to a surface flux of $10\,{\rm G}\,R_\odot^2$ to $10^4 {\rm G}\,R_\odot^2$, respectively. By
accretion, this field is strengthened as the star contracts.  So the
final $B_{\rm s}$ ranges from about $10^6$ to~$10^9\,$G. In the present work, we report the results for constant flux $10\,{\rm G}\,R_\odot^2$, although changing this only changes $B_s$ and thus does not change the results.

\section{Dependence on composition}
\label{sec:comp}

   To understand the effects of the composition of WD and the
    accreted material, we compare two models:
    \begin{itemize}
            \item Accretion on to a $1.02\,M_\odot$ CO WD, formed from a
          $8\,M_\odot$ MS \citep{IfmR_Binaries}.
        \item Accretion on to a $0.45\,M_\odot$ HeCO hybrid WD, formed
          from a $3\,M_\odot$ MS \citep{Cignition}.
    \end{itemize}

\begin{figure}[!htpb]
	\centering
	\includegraphics[scale=0.32]{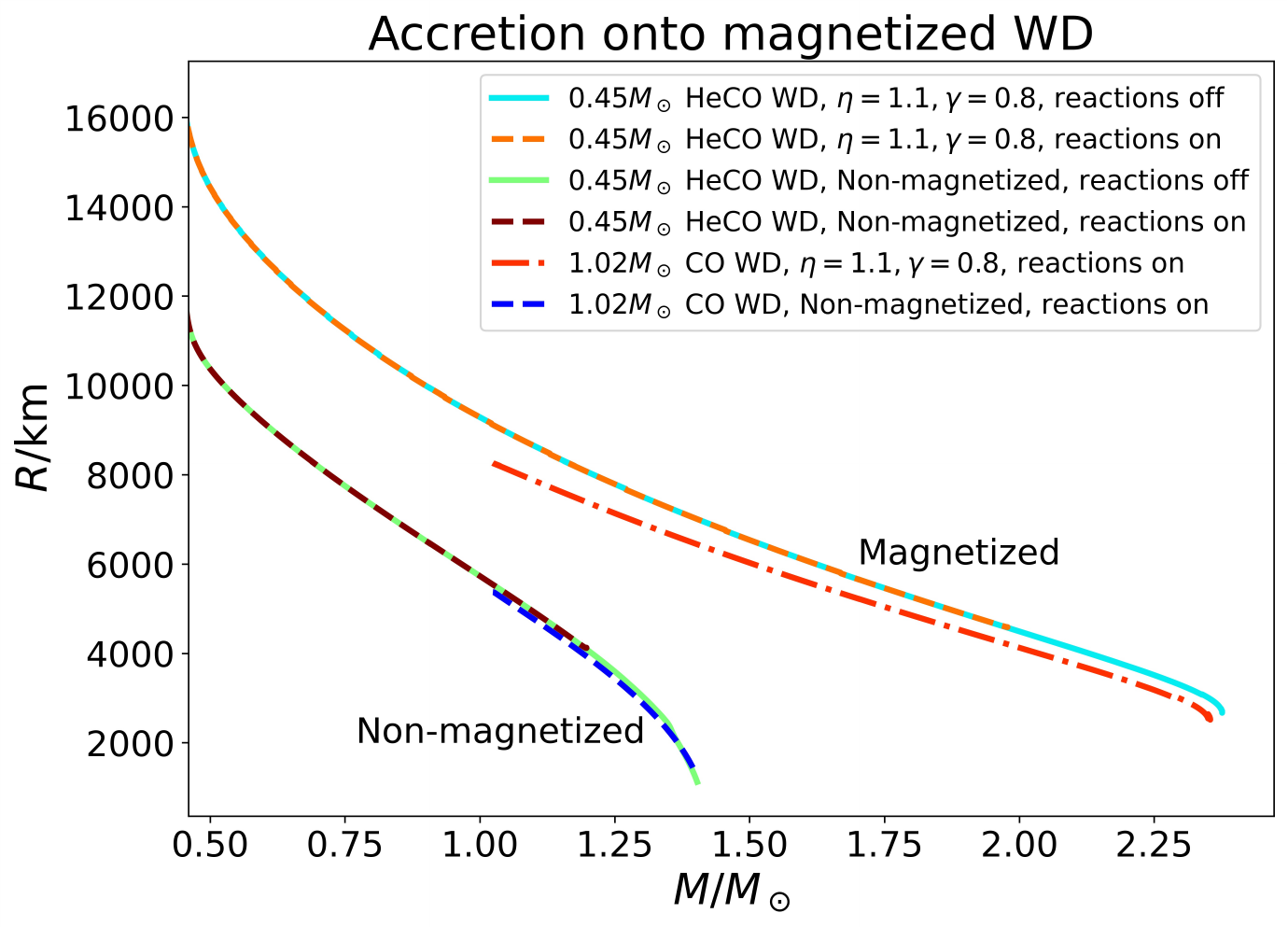}
	\caption{Modified $M-R$ relations for accreting non-magnetized and magnetized
$0.45M_\odot$ HeCO WDs and $1.02M_\odot$ CO WDs. In all magnetized cases, $\eta = 1.1, \gamma = 0.8$ ($\eta$ and $\gamma$ describe the magnetic field profile defined in the Appendix A).
    The end of each track denotes the maximum mass of a stable WD created. 
    In the HeCO cases, the triggering of unstable helium burning leads to WD being destabilized before mass limit is reached. We confirm this effect more clearly by comparing the cases between nuclear reactions enabled in the code (``reactions on'') and the cases when nuclear energy release does not go into the stellar structure equations (``reactions off"). 
          Core carbon ignition in the CO WDs occurs close
          to the maximum mass. }
    \label{HeCO_CO_acc}
\end{figure}

In both cases, the evolution to WD is described in the Appendix B.  
In the case of CO-WD with minimal helium confined only to the outermost layers, we consider $\dot{M}= 10^{-9} M_\odot {\rm yr}^{-1}$ CO-rich material. 
This is similar compositionally to the double degenerate (DD) scenario in which SNeIa are
triggered by merging two CO WDs \citep{dd_sn1a}.  Another possibility
is the core degenerate (CD) progenitor scenario where a WD merges with
the hot CO rich core of an AGB star \citep{core_degen}.  Both of these
models can lead to super-CM WDs before SNeIa as seen in Fig.~\ref{HeCO_CO_acc}, as the magnetization allows the
support of more mass. At higher accretion rates, $\dot{M} \gtrsim 10^{-6}\,M_\odot\,{\rm yr}^{-1}$ (see \citealt{Cignition}, for the possible increase and decrease of accretion rate depending on various possible mechanisms) as opposed to $\dot{M}\lesssim 10^{-7}\,M_\odot\,{\rm yr}^{-1}$ used in the current work, surface ignition of carbon that occurs before mass saturation/limit may however destabilize the star \citep{Cignition}. 

The case of hybrid HeCO WD is a bit subtle, in terms of unstable
nuclear burning and mass supply, both. This has around
$0.1\,M_\odot$ helium remaining. The presence of
significant helium in the hybrid WD triggers unstable nuclear burning in
the envelope and partly within the core when the accretion rate is
sufficiently high, e.g. $\gtrsim 10^{-9}\,M_\odot\,{\rm yr}^{-1}$. As a result, WD does not reach the saturated
maximum mass on the $M-R$ curve. 
This is true regardless of magnetic fields. The field profile and
strength only alter the mass at which the ignition takes place.  
Also, there is a uncertainty of helium mass-budget in donor to lead to $M_{\rm Ch}$ when the accretor itself is only $0.45M_\odot$. 
However, much more complicated processes than the route to the end point mentioned above and the exotic cases like triple star systems \citep{tripleWD} are not ruled out. Nevertheless, for a non-magnetic HeCO WD with significant helium, ignition occurs below $M_{\rm Ch}$,
which prevents further evolution with the STARS code. 
Indeed, the Chandrasekhar and other
saturation mass limits are ideal maximum masses for WDs.  
In reality, other effects, such as residual helium burning (as in this case),
pycnonuclear reactions \citep{chamel} etc., can destabilize the WD
before it reaches its ideal maximum limit. This is the case
for any SNeIa.  Nevertheless, depending on the magnetic profile and
strength, B-WDs can reach super-CMs before nuclear
burning destabilizes them, if there is adequate mass in donor(s).  If we, however, switch off helium burning in the code
before it ignites, then the WD follows the $M-R$ relation to the
maximum mass.  Moreover, when negligible helium is present, its burning is not violent so that both magnetic and non-magnetic WDs accrete
to their corresponding maximum masses following the appropriate $M-R$ relations. This confirms the destabilization is due to helium burning.

Fig.~\ref{HeCO_CO_acc} illustrates this evolution and
confirms that B-WDs can reach significantly super-CMs
before any nuclear burning ignites. For HeCO WDs, we consider
$\dot{M} = 10^{-9} M_\odot {\rm yr}^{-1}$. Such a low $\dot{M}$
($\lesssim 10^{-7} M_\odot {\rm yr}^{-1}$) is
considered in accordance with STARS capability of mass loading which, however, does not include any physics of cataclysmic variables. Hence this may additionally trigger helium flashes at the surface, leading to novae like eruptions \citep{heflash,bowang_nova}. The realistic accretion rates of these scenarios, calculated with the consideration of the physics of cataclysmic variables, novae etc., may be higher \citep{Cignition,flash}. Further, for $B_s > 10^{6} \ G$, the accretion may be channeled through a column rather than the spherical mass addition we have considered here, leading to surface effects. We leave these studies to future work with specialized codes.

Accretion of helium-rich material on to WDs is popular in some models of
SNIa progenitors \citep{he_acc1}.  In particular, this can produce the
under-luminous class of SNIa, namely SNIax \citep{Iax} as well as SN.Ia \citep{dotIa}. Other faint transient classes have also been associated
with the helium accretion onto hybrid HeCO WDs \citep{HeCO_WD}
as discussed in this section and hence, our results are useful for comparison
with observations. Additionally, the accretion of helium-rich material at an ``effective rate" is, on a simulation basis, equivalent to a long-term study of accreting hydrogen-rich (solar composition) material. In this case, one needs to account for hydrogen flashes as well, which may work against the star gaining mass~\citep{flash}. However,
it is possible that an accreting WD in reality could evolve through these nuclear flashes and remain stable as shown by other hydrodynamical models \citep{flash}. 
It has been also confirmed by previous numerical simulations that the mass-budget of donor-accretor pairs \citep{flash} to form a super-CM B-WD is adequate.

\section{Observations of expanded radii: combining magnetic and finite temperature effects}
The finite temperature of WDs can have important implications on their
observables, in particular, their radii. As shown in earlier works, e.g. \citealt{abhay,Tdep}, this finite temperature effect alone can lead to WDs deviating from the $M-R$
relation, similarly to the effects of magnetic fields. Further, toroidal field dominated WDs have radii larger than the Chandrasekhar predictions, while poloidally dominated fields tend to show smaller radii \citep{deb}. We find the same trend in our simulations, though we focus on toroidally dominated WDs based on stability arguments \citep{BR2009}. We investigate
the properties of magnetized and non-magnetized WDs cooled to
different surface temperature ($T_e$). Here, the WDs of various masses (both magnetized and non-magnetized), formed by accretion, follow their respective cooling track after their formation. WDs of a given temperature are then grouped together to get $T_e$ specific $M-R$ relations.

\begin{figure}[!htb]
     \centering

         {\centering
         \includegraphics[width=0.45\textwidth]{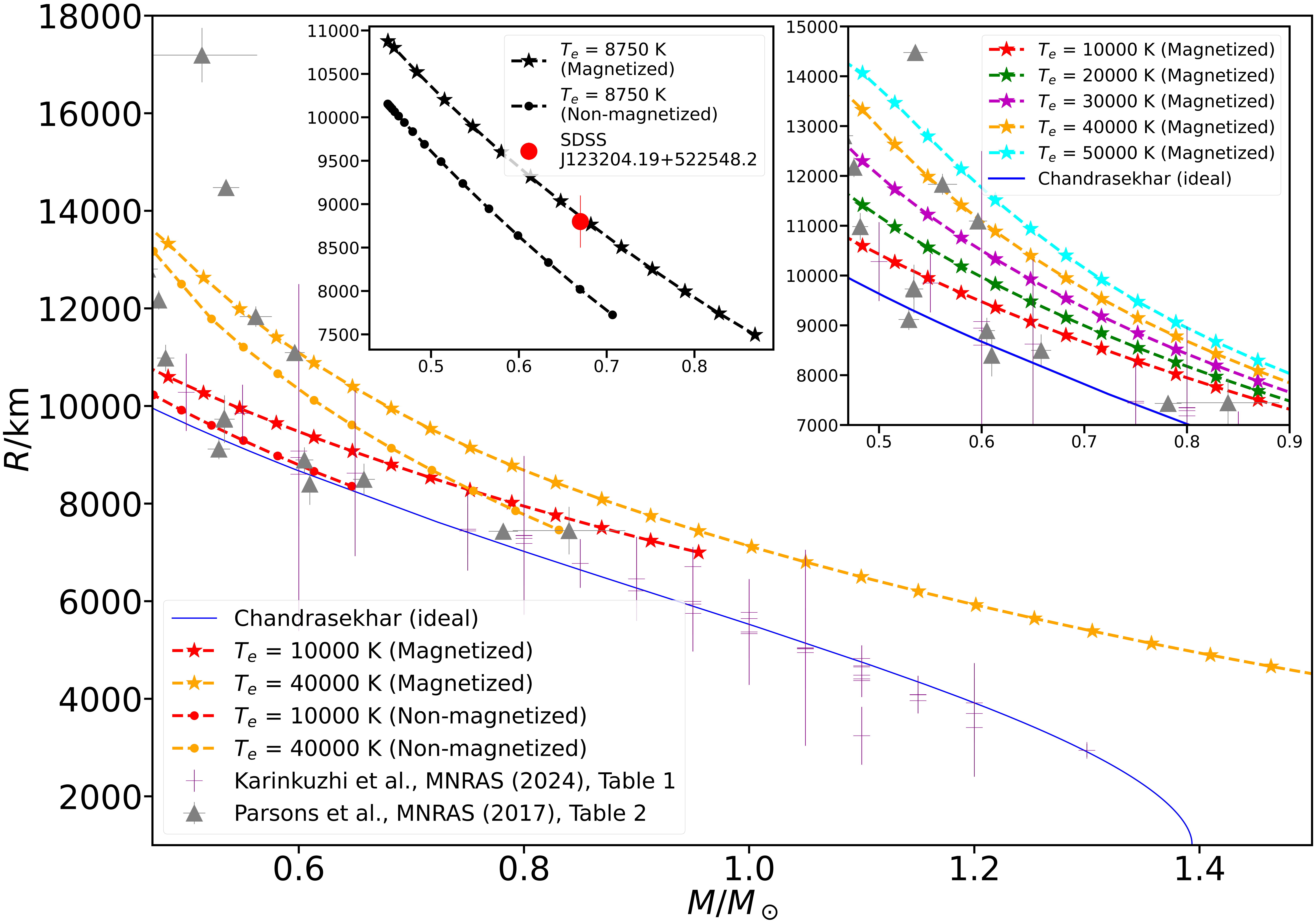}

     \vskip\baselineskip

            \includegraphics[width=0.45\textwidth]{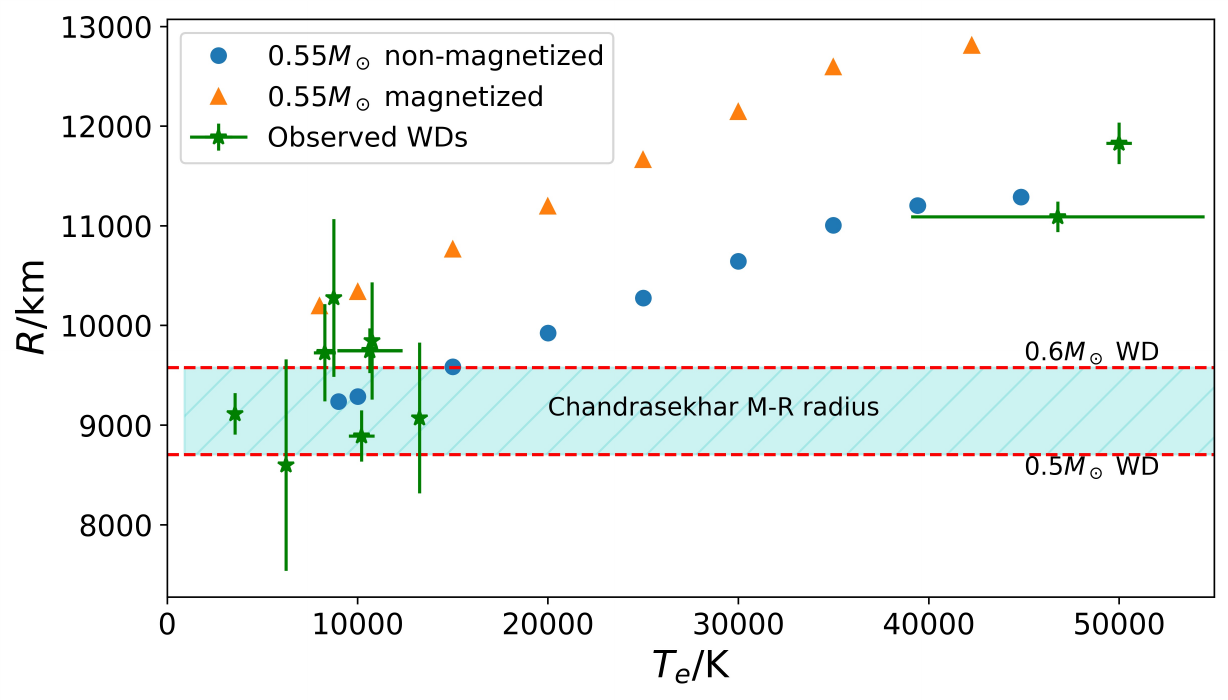}}

\caption{\textit{Upper}: effect of temperature and/or magnetic field on the $M-R$ relation. 
The  WDs formed by accretion on to a $0.45M_\odot$ HeCO WD are further cooled to various $T_e$ and the $M-R$ relations for a specific $T_e$ are shown. Non-magnetized cases are shown with dashed-line with filled circle, and magnetized cases with dashed-line with asterisk. The ideal Chandrasekhar's case is also shown by a solid line. Low-mass
          outliers \citep{parsons, drisya} mostly fall within the extent of
          these curves. The rightmost inset shows many more magnetized cases with various $T_e$ for completeness. The leftmost inset shows the particular case of SDSS J123204.19+522548.2 \citep[$T_e = 8750$ K: ][]{drisya}, confirming magnetized WD model suffices to explain the data. \textit{Lower}: the dependence of radius on $T_e$ for both magnetized (filled triangle) and non-magnetized (filled circle) cases of $M=0.55M_\odot$. Observed outliers shown therein are for $0.5\lesssim M/M_\odot\lesssim 0.6$, taken from the ones displayed in upper panel. The ideal zero-temperature result is shown by a zone bounded by dashed lines. In the magnetic cases, the field is specified by the profile parameters $\eta = 1.3, \gamma = 0.9$ (defined in Appendix A).}
    \label{TandB}
\end{figure}

Fig.~\ref{TandB} illustrates the combined effects of
temperatures and magnetic fields and observed $M-R$ points of $0.5\lesssim M/M_\odot\lesssim 0.6$.
Comparing the $M-R$ curves between
magnetized and non-magnetized WDs at particular respective $T_e$,
we see that the magnetic field increases the deviation from
Chandrasekhar's $M-R$ relation, see particularly lower panel emphasizing deviation in $R$. Thus any observed deviation of $R$
for a given $M$ can be due to a combination of both magnetic and temperature effects, as opposed to the temperature effect alone. Other possible effects, e.g. the presence of a hydrogen envelope, He core as opposed to CO core, could also add to it \citep{parsons}.
Some outlying WDs, e.g. SDSS J123204.19+522548.2 with $T_e = 8750$ K, with independently observed masses and radii
\citep{parsons,drisya}, require both effects to explain their large
deviations for observed $T_e$, as shown in the leftmost inset of the upper panel of Fig. 4. In this case, the radius obtained by considering both magnetic field and finite $T_e$ ($0.7\%$ error) is an order of magnitude better fit to the observed radius as opposed to that calculated from just the finite temperature effect ($7\%$ error). See Appendix for other details. We leave a more systematic study that incorporates uncertainties/degeneracies related to observational errors, different magnetic field profiles and the other effects mentioned above to future work.

\section{Discussion}
\label{sec:discussion}

The idea of violation of $M_{\rm Ch}$, sacrosanct for
non-magnetized, non-rotating cold WDs, and the existence of
super-CM WDs, is intriguing, even without their link to
cosmology. While highly magnetized WDs have been proposed before,
these had not been established by any detailed stellar evolution modeling. 
As a result, time to time skepticism appeared in the literature with respect to their magnetohydrodynamical and nuclear stability, mass, underlying bound of magnetic field and geometry (e.g.   \citealt{coelho,nitya,bera,chatterjee}), which however were also shown to be based on erroneous calculations and assumptions (e.g. \citealt{dasmukhoMPLA,dasmukhoPRDReply,sathya,abhay,BMVirial}). Nevertheless,
only by
full stellar evolution, with stars passing through complicated 
phases, including nuclear and (magneto)hydrodynamical instabilities,
mass loss and cooling, with embedded magnetic flux and full comparison
with observations, can we refute all the doubts with utmost confidence and confirm their existence.  Our exploration brings
them closer to 
reality with implications from astrophysics to cosmology.

Our work has the following novelties: \begin{enumerate}
    \item modeling the time evolution of magnetized MSs to WDs with a detailed, stable, stellar evolutionary path forming significantly super-CM WDs, with a modified $M-R$ relation,
    \item explaining unusually large sub-CM WDs deviated from Chandrasekhar's $M-R$ relation by magnetic field and temperature,
    \item exploring modified H-R diagram in the presence of magnetic field, leading to both the above-mentioned unusual objects,
    \item suggesting while “Chandrasekhar limit” may be sacrosanct, its value severely depends on the underlying physics, here varied effects of magnetic field, determining a series of $M-R$ curves with respective mass limits.
\end{enumerate} 

The nature of SNeIa as standard candles is one of the probes of dark
energy.  With them, the accelerating expansion of the Universe was first
found \citep{darkenergy1,darkenergy2} and the Hubble
constant tied to greater precision \citep{branch1992}.  Either violation or non-uniqueness of $M_{\rm Ch}$
questions the standard candle nature of SNeIa, bringing with it many
important consequences for cosmology, which may have  an impact on Hubble tension \citep{trg_hubble}.

Importantly, for some
stellar compositions, mass loss/accretion, and magnetic fields, WDs follow a stable evolutionary path to the 
point of mass limit.
This evolutionary path, however, ceases for arbitrary parameters, e.g., extremely high central fields and steep magnetic field profiles result in incomplete evolution. As we have elucidated here, other effects such as those based on (odd) composition may end up destabilizing the star independent of the magnetic field.

Although there is an apparent caveat due to one-dimensional approximation, introducing effects like rotation do not affect the qualitative results. Indeed, WD's Keplerian frequency is very small compared to a neutron star. However, the future aim should be to consider the \textit{explicit back-reaction} of the magnetic field on stellar properties like convection, mixing length etc., which requires explicit Maxwell solutions along with time-dependent stellar evolution, which is beyond the scope of the evolutionary codes available today. As the magnetic field remains dormant for most of the evolutionary history of the star, we expect that the results will not be affected too significantly.

Nevertheless, we confirm
the possibility of very luminous, with super-CM,
progenitor of SNeIa, for viable field geometries.  
Additionally, the combined effects
of magnetic fields and temperature can explain many observed
sub-CM WDs which do not conform with Chandrasekhar $M-R$ curve, hence apparently are outliers. 

B-WDs can further be linked to other unusual observations \citep{bm_sgraxp}. There have been several detections of slowly to very slowly rotating
pulsars of period $P$ as high as $1000\,$s \citep[e.g. GPM~J1839$-$10: ][]{gpm},  
which are not confirmed neutron star (NSs). Their high $P$s naturally 
suggest WDs. Even if not yet, these are potential candidate super-CM WDs after
further accretion \citep{arsco_bm}. Also the
magnetic moon-size WD ZTF J1901+1458, with mass $1.35 M_\odot$, $B_s \sim 8\times 10^8$ and $T_e\sim 46000$ K \citep{Caiazzo-Nature}, could be in the $M-R$ curve of  centrally poloidally dominated B-WDs, unlike our toroidally dominated cases. This is indicative from previous work \citep{sathya,deb2} and should be explored in detail.
 
Future observation and modelling will further and more robustly
illuminate these issues.

\begin{acknowledgements}
    The authors thank A. R. Rao of TIFR, for many illuminating discussions throughout the project. They also gratefully thank the reviewer for a very encouraging report with constructive comments and suggestions, which have helped to prepare the final version.  BM thanks
the Institute of Astronomy, University of Cambridge, for hospitality where the last part of the work was done. ZZ acknowledges the Prime Minister's Research Fellows (PMRF)
scheme, with Ref. No. TF/PMRF-22-7307, for providing fellowship. AJH thanks the Czech Grant Agency (GAČR) for his funding under the grant number 21-16583M.  CAT thanks Churchill College for
his fellowship. BM acknowledges the partial support from a project funded by SERB, India, with Ref. No.
CRG/2022/003460.
\end{acknowledgements}

\begin{contribution}
ZZ performed and analyzed the stellar evolution simulations, and wrote and edited the manuscript. BM analyzed the data, supervised the project and wrote and edited the manuscript. AK performed and analyzed the stellar evolution simulations. AS, AJH and CAT assisted with the stellar evolution simulations. PB verified the underlying physics and edited the manuscript. CAT proposed and verified the idea of accreting WD evolution, verified the underlying physics of simulated results and edited the manuscript.


\end{contribution}

%

\software{Cambridge STARS \citep{eggleton, eggleton2, pols}, Matplotlib \citep{matplotlib}, NumPy \citep{numpy}.}


\appendix

\section{Simulation set-up and formalism}
\label{sec: simulation}

The Cambridge STARS stellar evolution code has successfully been used to model a wide range of stars in
a wide range of evolutionary phases over the decades.  Some progress
had been made to include the effects of strong magnetic fields on WDs
\citep{mukul}.  We have modified STARS to include such fields.  Rather
than simultaneously solving Maxwell's equations and magnetostatic
balance in a Newtonian model, we introduce, following previous practice
\citep{mukul}, a density-dependent magnetic field profile
\citep{bandyopadhyay} that fixes the magnitude of magnetic field for a
given density, coupled with the mass and radius of the star.  This is
a valid assumption as long as the star is approximately spherical
\citep{deb,deb2}. Throughout, we have the following assumptions and
considerations.

\begin{itemize}
    \item The star is spherical and one-dimensional.
    \item The star is non-rotating.
    \item We solve for the magnetohydrostatic balance equations by modifying the
      pressure and density to account for the effects of the magnetic
      field
\end{itemize}

The stellar structure equations used are
\begin{equation}
    \frac{d\log_e P}{dm} = -\frac{Gm}{4\pi r^4P},
\end{equation}

\begin{equation}
    \frac{d\log T}{dm} = \frac{d\log P}{dm}\nabla,
\end{equation}

\begin{equation}
    \frac{d \log_e r}{dm} = \frac{1}{4\pi r^3m},
\end{equation}
and
\begin{equation}
    \frac{dL}{dm} = \epsilon - \epsilon_\nu - \frac{Du}{Dt} + \frac{P}{\rho^2}\frac{D\rho}{Dt},
\end{equation}
where $m$ is the mass at radius $r$ in a star at time $t$, and $T$,
$P$, $\epsilon$, $\epsilon_\nu$ and $\rho$ are the corresponding
temperature, pressure, energy generation rate by nuclear reactions,
neutrino energy loss rate and density respectively.  The logarithmic
temperature gradient with pressure, $\nabla$, depends on whether the star
is locally convective or radiative.  We solve for a network
of 21 nuclear reactions that are important for stellar structure.

We incorporate magnetic fields with an approximate mixed field
geometry by modifying the pressure. We add contributions from both magnetic pressure $P_B=B^2/2\mu_0$, where $\mu_0$
is the permeability of free space, and magnetic tension to the
matter pressure ($P_m$).  We consider the innermost regions of
the star to have a toroidal field with total pressure $P=P_m + P_B$,
mimicking field lines orthogonal to the radial direction.  The outer
layers of the star ($10-30\%$ or so in radius) have a poloidal field
contribution with total pressure by $P = P_m - P_B$, mimicking radial
field \citep{deb,deb2,zz}.  This is a one-dimensional approximation of the mixed field
geometry\footnote{This follows from the magnetization tensor, i.e. $M_{ij}~=~(B^2/8\pi)\delta_{ij}~-~B_iB_j/4\pi$, where $B_i$ are the components of the magnetic field in Minkowski space. In the case of radially oriented fields, $\textbf{B} = (B_r, 0, 0)$ and for transverse oriented fields, $\textbf{B} = (0, B_\theta, B_\phi)$.}. This is stable in the interior \citep{BR2009}
because stability relies on the ratio of the poloidal field energy to
the sum of poloidal and toroidal field energy being in the
range from about $10^{-3}$ to $0.8$ so that toroidally dominated
geometries are more probable \citep[{see also:}][]{tout_mostmagnstars}. In the current work, we consider the cases with poloidal field confined to the outer $10\%$ of the star by radius. The maximum mass $M_{\mathrm{max}}$ increases by a maximum of $7\%$ when the poloidal region decreases from $30\%$ to $10\%$.  

The above cases deviate the star marginally from spherical symmetry, with ellipticity estimated as \citep{ellipticity,zz_csc}
\begin{equation}
    \epsilon = \pi/I_0 \int_V dr\ \delta\rho \ r^4,
    \label{el}
\end{equation}
where $\delta\rho = B^2/8\pi c^2$ is the density perturbation arising due to the magnetic field and $I_0$ is the unperturbed moment of inertia.

Our density-dependent model magnetic field
profile \citep{bandyopadhyay} is
\begin{equation}
\label{prof}
B(\rho) = B_{\rm s} + B_0\left[1 - \exp\left\{-\eta{\left(\frac{\rho}{\rho_0}\right)^\gamma}\right\}\right],
\end{equation}
where $B_{\rm s}$, $B_0$, $\eta$ and $\gamma$ are
model parameters that determine the strength and profile of the
magnetic field from the center of the star to its surface. By equations \eqref{prof} and \eqref{el}, $\epsilon \lesssim 10^{-7}$ for the WDs formed at the end of stellar evolution with dormant magnetic fields, and $\lesssim 10^{-5}$ for the massive super-Chandrasekhar B-WDs under consideration. This justifies the use of a one-dimensional code.

Due to the density dependence of $B$, as the star contracts leading to the denser parts of its evolution, its magnetic field is also strengthened. This is an approximation of flux freezing. Stars of higher densities and, hence, smaller radii have higher magnetic fields throughout. In the present work, we fix $B_0 = 10^{14}\,$G and $\rho_0 = 10^9\,\rm g\,cm^{-3}$.
We further fix $B_{\rm s}$ according to explicit ideal magnetohydrodynamic (MHD)
flux freezing approximation such that $B_{\rm s}R^2$ is constant. We consider surface fields of $10\,$G when $R = R_\odot$. As the central density ($\rho_c$) for the MSs is $\sim 1-10$ g cm$^{-3}$, the central magnetic field from equation \eqref{prof} is $B_{\rm c}\approx 10^6 - 10^7\,$G. We evolve this kind of magnetized MS. However, $B_{\rm s}$ does not significantly
affect the maximum mass of the WD.   The extra mass is largely stabilized
by the central field. Nevertheless, $B_{\rm s}$ is important to
compare models with observations.

\section{Formation of magnetized WDs} 
To model magnetized WDs or B-WDs, we start with zero-age main-sequence
(ZAMS) models.  The composition of the WD is determined by the initial
ZAMS mass and the rate of mass loss. CO WDs
typically derive from stars with ZAMS masses within the range $3$
to~$8\,M_\odot$ and when the envelope is lost on the AGB but before
carbon ignition.  He WDs are from MSs of lower mass when
the envelope is removed on the red giant branch (RGB), before helium ignition.
Oxygen--neon (ONe) WDs can be formed from super-AGB stars that lose
their mass on the AGB but after core carbon burning. We do not include the effects of convective overshooting in the present work. It may affect the core compositions around $8M_\odot$, but does not affect the eventual mass limit(s) of the 
B-WDs.

We evolve these MSs to the end of their nuclear burning. The
final state is the result of a competition between nuclear burning and
mass loss driving the evolution.  For example, CO WDs evolve past core He
burning on to AGB when the star has an appreciable CO core,
typically exceeding $0.5\,M_\odot$.  Hydrogen and helium shell burning
at the edge of the shrinking CO core grows the core and increases the
star's luminosity.  The envelope expands and is stripped off in a
stellar wind at a high rate (typically with $\gtrsim10^{-6}\,M_\odot\,{\rm yr}^{-1}$). The chosen AGB mass loss rates fall within the uncertainties discussed in the
literature \citep{agb_rate}. We use the code's spherical mass loss control, to mimic AGB mass losses. High mass loss rates can also mimic binary formation processes~\citep{Cignition,IfmR_Binaries}. 
 The hydrogen shell is then
extinguished leaving a hot WD that contracts and cools to a 99\%
degenerate WD. Throughout this process, we have magnetic field imposed by means of the model magnetic field profile, described earlier. The magnetic field grows with density as the star moves to denser parts of its evolution. 

The masses of the final WDs range from $0.45$ to~$1.02\,M_\odot$, evolved from MSs of masses from $3$ to~$8\,M_\odot$. We consider different values of $\eta$ and $\gamma$ in the present work, corresponding to different magnetic field profiles. The values of $\eta$ and $\gamma$ are chosen such that we obtain complete evolution, without destabilization. Very steep magnetic field profiles can lead to destabilization and incomplete evolution. The values chosen here are not the only possibilities - see previous work \citep{zz} for how the $M-R$ relation and other stellar properties change with change of $\eta$ and $\gamma$. The
cooled WDs then have $B_{\rm s}$ in the range $(1$ to~$2.5)\times 10^5\,$G
and central fields $0.5\lesssim B_{\rm cent}/10^{12}\,G \lesssim 3$. They further grow by accretion.

\section{Consistency with observed magnetic fields}
    
Only $B_{\rm s}$ of WDs can be measured observationally. 
The strongest magnetic fields in MSs are found in A and~B stars
with $B_{\rm s} \approx 10^3\,$G.  These are consistent with our chosen
fields in the range of $1$ to~$10^3\,$G \citep{tout_mostmagnstars}.
However, even stronger $B_{\rm s}$ cannot be ruled out because observational
data become difficult to interpret  \citep{drisya}.

\begin{figure}[!htpb]
	\centering
	\includegraphics[scale=0.5]{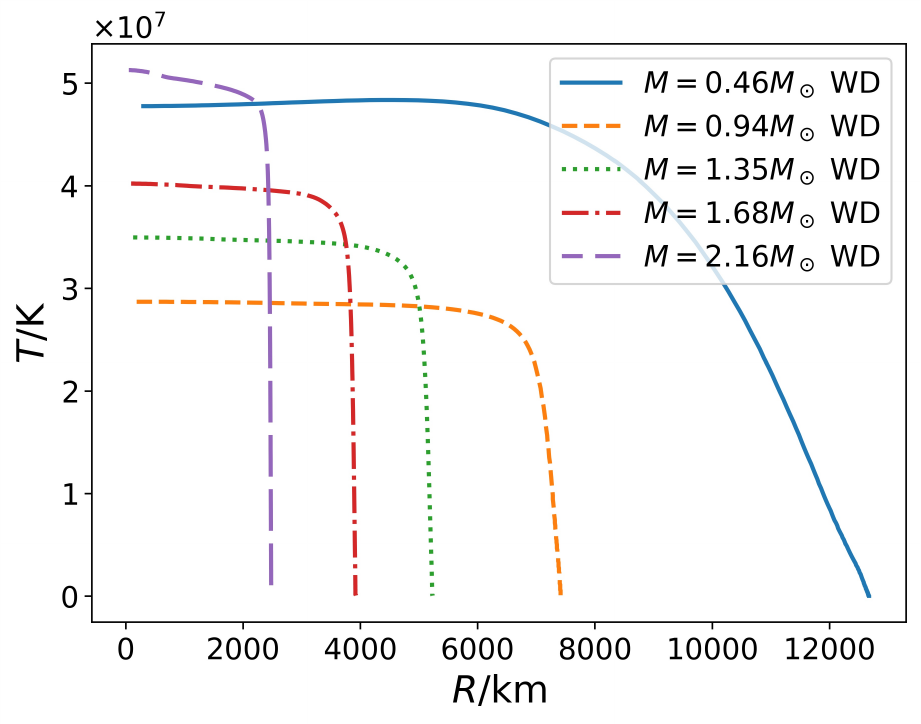}
	\caption{Profiles showing change of temperature from centre to surface for a few different magnetized WDs of different mass. Here, magnetic field profile is given by $\eta = 1.1, \gamma = 0.9.$
   }
    \label{temp}
\end{figure}

We simplify the evolution of the field due to the evolution to the WD
by assuming $B_{\rm s}$ grows from its MS strength by flux
freezing. Fig. \ref{temp} shows that the temperature inside WDs remains 
almost constant throughout except very close to the surface, particularly
for smaller WDs. This argues for high conductivity therein, supporting our frozen-flux
assumption. Our WDs, including B-WDs with fields enhanced through mass accretion, then have $10^4\lesssim B_{\rm s}/G\lesssim
10^{9}$, which is consistent with the fields of observed magnetic WDs
\citep{ferrario}.  Once again these fields could be stronger for WDs
evolved from more strongly magnetized MSs.  Alternatively, the
magnetic flux may grow by a dynamo mechanism \citep{quentin2018} 
but we do not explore this here.

\section{Magnetic field decay}
Processes that include the Ohmic decay and Hall effect can lead
magnetic field to decay in WDs on long time-scales
\citep{mukul,goldreich,heylkulkarni}.  We investigate the effects of
such decay, if any, with a formalism developed earlier
\citep{mukul,goldreich,heylkulkarni} such that
\begin{equation}
    \frac{dB}{dt} = -B\left(\frac{1}{t_{\rm Ohm}}+\frac{1}{t_{\rm Hall}}\right),
\end{equation}
where $B$ is the magnitude of magnetic field, given by
equation~(\ref{prof}).  The characteristic
time-scales \citep{mukul} are given by
\begin{equation}
    t_{\rm Ohm} = 7\times 10^{10}\,\rho_{c,6}^{1/3}R_4^{1/2}\frac{\rho_{\rm av}}{\rho}\,{\rm yr}
\end{equation}
and
\begin{equation}
    t_{\rm Hall} = 5\times 10^{10}\,l_8^2 B_{0,14}^{-1} T_{\rm c,7}^2\rho_{\rm c,10}\,{\rm yr},
\end{equation}
where $\rho_{{\rm c},n} = \rho_{\rm c}/10^n\,\rm g\,cm^{-3}$,
$\rho_{\rm c}$ is the central density of the WD, $\rho_{av}$ is its
mean density, $R_{4} = R/10^4\,$km, $T_{c,7} = T_{\rm c}/10^7\,$K,
$T_{\rm c}$ is the central temperature, $B_{0,14} = B_0/10^{14}\,$G
and $l = l_8\times 10^8\,$cm is characteristic length scale of the
flux loops through the outer core of the WD.

We find that it takes about 5\,Gyr for the fields under consideration
to decay and thence for the WD radius to shrink significantly.  However,
our accretion lasts at most $\sim1$\,Gyr.  So we ignore field decay.


\bibliography{biblio}{}

\begin{thebibliography}{}
\expandafter\ifx\csname natexlab\endcsname\relax\def\natexlab#1{#1}\fi
\providecommand{\url}[1]{\href{#1}{#1}}
\providecommand{\dodoi}[1]{doi:~\href{http://doi.org/#1}{\nolinkurl{#1}}}
\providecommand{\doeprint}[1]{\href{http://ascl.net/#1}{\nolinkurl{http://ascl.net/#1}}}
\providecommand{\doarXiv}[1]{\href{https://arxiv.org/abs/#1}{\nolinkurl{https://arxiv.org/abs/#1}}}

\bibitem[{D. {Adam}(1986){Adam}}]{adam}
{Adam}, D. 1986, \bibinfo{title}{{Models of magnetic white dwarfs.},} \aap,
  160, 95

\bibitem[{D. {Bandyopadhyay} {et~al.}(1997){Bandyopadhyay}, {Chakrabarty}, \&
  {Pal}}]{bandyopadhyay}
{Bandyopadhyay}, D., {Chakrabarty}, S., \& {Pal}, S. 1997,
  \bibinfo{title}{{Quantizing Magnetic Field and Quark-Hadron Phase Transition
  in a Neutron Star},} \prl, 79, 2176, \dodoi{10.1103/PhysRevLett.79.2176}

\bibitem[{P. Bera \& D. Bhattacharya(2017)Bera \& Bhattacharya}]{bera}
Bera, P., \& Bhattacharya, D. 2017, \bibinfo{title}{{A perturbation study of
  axisymmetric strongly magnetic degenerate stars : the case of
  super-Chandrasekhar white dwarfs},} Mon. Not. Roy. Astron. Soc., 465, 4026,
  \dodoi{10.1093/mnras/stw2979}

\bibitem[{O. {Bertolami} \& H. {Mariji}(2016){Bertolami} \&
  {Mariji}}]{ungravity}
{Bertolami}, O., \& {Mariji}, H. 2016, \bibinfo{title}{{White dwarfs in an
  ungravity-inspired model},} \prd, 93, 104046,
  \dodoi{10.1103/PhysRevD.93.104046}

\bibitem[{M. {Bhattacharya} {et~al.}(2022){Bhattacharya}, {Hackett}, {Gupta},
  {Tout}, \& {Mukhopadhyay}}]{mukul}
{Bhattacharya}, M., {Hackett}, A.~J., {Gupta}, A., {Tout}, C.~A., \&
  {Mukhopadhyay}, B. 2022, \bibinfo{title}{{Evolution of Highly Magnetic White
  Dwarfs by Field Decay and Cooling: Theory and Simulations},} \apj, 925, 133,
  \dodoi{10.3847/1538-4357/ac450b}

\bibitem[{L. {Bildsten} {et~al.}(2007){Bildsten}, {Shen}, {Weinberg}, \&
  {Nelemans}}]{dotIa}
{Bildsten}, L., {Shen}, K.~J., {Weinberg}, N.~N., \& {Nelemans}, G. 2007,
  \bibinfo{title}{{Faint Thermonuclear Supernovae from AM Canum Venaticorum
  Binaries},} \apjl, 662, L95, \dodoi{10.1086/519489}

\bibitem[{J. {Braithwaite}(2009){Braithwaite}}]{BR2009}
{Braithwaite}, J. 2009, \bibinfo{title}{{Axisymmetric magnetic fields in stars:
  relative strengths of poloidal and toroidal components},} \mnras, 397, 763,
  \dodoi{10.1111/j.1365-2966.2008.14034.x}

\bibitem[{D. {Branch}(1992){Branch}}]{branch1992}
{Branch}, D. 1992, \bibinfo{title}{{The Hubble Constant from Nickel
  Radioactivity in Type IA Supernovae},} \apj, 392, 35, \dodoi{10.1086/171401}

\bibitem[{I. Caiazzo {et~al.}(2021)Caiazzo {et~al.}}]{Caiazzo-Nature}
Caiazzo, I., {et~al.} 2021, \bibinfo{title}{{A highly magnetised and rapidly
  rotating white dwarf as small as the Moon},} Nature, 595, 39,
  \dodoi{10.1038/s41586-021-03799-3}

\bibitem[{N. {Chamel} {et~al.}(2013){Chamel}, {Fantina}, \& {Davis}}]{chamel}
{Chamel}, N., {Fantina}, A.~F., \& {Davis}, P.~J. 2013,
  \bibinfo{title}{{Stability of super-Chandrasekhar magnetic white dwarfs},}
  \prd, 88, 081301, \dodoi{10.1103/PhysRevD.88.081301}

\bibitem[{S. Chandrasekhar(1935)Chandrasekhar}]{chandralimit}
Chandrasekhar, S. 1935, \bibinfo{title}{{The highly collapsed configurations of
  a stellar mass (Second paper)},} Mon. Not. Roy. Astron. Soc., 95, 207,
  \dodoi{10.1093/mnras/95.3.207}

\bibitem[{D. {Chatterjee} {et~al.}(2017){Chatterjee}, {Fantina}, {Chamel},
  {Novak}, \& {Oertel}}]{chatterjee}
{Chatterjee}, D., {Fantina}, A.~F., {Chamel}, N., {Novak}, J., \& {Oertel}, M.
  2017, \bibinfo{title}{{On the maximum mass of magnetized white dwarfs},}
  \mnras, 469, 95, \dodoi{10.1093/mnras/stx781}

\bibitem[{J.~G. Coelho {et~al.}(2014)Coelho, Marinho, Malheiro, Negreiros,
  Rueda, Ruffini, \& C{\'a}ceres}]{coelho}
Coelho, J.~G., Marinho, R.~M., Malheiro, M., {et~al.} 2014,
  \bibinfo{title}{{Dynamical instability of white dwarfs and breaking of
  spherical symmetry under the presence of extreme magnetic fields},}
  Astrophys. J., 794, 86, \dodoi{10.1088/0004-637X/794/1/86}

\bibitem[{N.~R. {Crumpler} {et~al.}(2024){Crumpler}, {Chandra}, {Zakamska},
  {Adamane Pallathadka}, {Arseneau}, {Gentile Fusillo}, {Hermes}, {Badenes},
  {Chakraborty}, {G{\"a}nsicke}, \& {Schmidt}}]{Tdep}
{Crumpler}, N.~R., {Chandra}, V., {Zakamska}, N.~L., {et~al.} 2024,
  \bibinfo{title}{{Detection of the Temperature Dependence of the White Dwarf
  Mass{\textendash}Radius Relation with Gravitational Redshifts},} \apj, 977,
  237, \dodoi{10.3847/1538-4357/ad8ddc}

\bibitem[{M. {Das} {et~al.}(2025){Das}, {Mukhopadhyay}, \&
  {Bulik}}]{mayusree_wd}
{Das}, M., {Mukhopadhyay}, B., \& {Bulik}, T. 2025, \bibinfo{title}{{Continuous
  Gravitational Waves from Magnetized White Dwarfs: Quantifying the Detection
  Plausibility by LISA},} \apj, 995, 107, \dodoi{10.3847/1538-4357/ae1732}

\bibitem[{M. Das {et~al.}(2025)Das, Sedrakian, \& Mukhopadhyay}]{mayusree2}
Das, M., Sedrakian, A., \& Mukhopadhyay, B. 2025, \bibinfo{title}{{Topology of
  the Superconducting Heart of Neutron Stars: Effects of Microphysics and
  Gravitational-Wave Signatures},} \doarXiv{2508.18363}

\bibitem[{U. {Das} \& B. {Mukhopadhyay}(2013){Das} \& {Mukhopadhyay}}]{udas}
{Das}, U., \& {Mukhopadhyay}, B. 2013, \bibinfo{title}{{New Mass Limit for
  White Dwarfs: Super-Chandrasekhar Type Ia Supernova as a New Standard
  Candle},} \prl, 110, 071102, \dodoi{10.1103/PhysRevLett.110.071102}

\bibitem[{U. Das \& B. Mukhopadhyay(2014)Das \& Mukhopadhyay}]{dasmukhoMPLA}
Das, U., \& Mukhopadhyay, B. 2014, \bibinfo{title}{{Revisiting some physics
  issues related to the new mass limit for magnetized white dwarfs},} Mod.
  Phys. Lett. A, 29, 1450035, \dodoi{10.1142/S0217732314500357}

\bibitem[{U. Das \& B. Mukhopadhyay(2015{\natexlab{a}})Das \&
  Mukhopadhyay}]{udas_modified}
Das, U., \& Mukhopadhyay, B. 2015{\natexlab{a}}, \bibinfo{title}{{Modified
  Einstein's gravity as a possible missing link between sub- and
  super-Chandrasekhar type Ia supernovae},} JCAP, 05, 045,
  \dodoi{10.1088/1475-7516/2015/05/045}

\bibitem[{U. Das \& B. Mukhopadhyay(2015{\natexlab{b}})Das \&
  Mukhopadhyay}]{dasmukhoPRDReply}
Das, U., \& Mukhopadhyay, B. 2015{\natexlab{b}}, \bibinfo{title}{{Reply to
  ''Comment on Strongly magnetized cold degenerate electron gas: Mass-radius
  relation of the magnetized white dwarf''},} Phys. Rev. D, 91, 028302,
  \dodoi{10.1103/PhysRevD.91.028302}

\bibitem[{D. {Deb} {et~al.}(2021){Deb}, {Mukhopadhyay}, \& {Weber}}]{deb2}
{Deb}, D., {Mukhopadhyay}, B., \& {Weber}, F. 2021, \bibinfo{title}{{Effects of
  Anisotropy on Strongly Magnetized Neutron and Strange Quark Stars in General
  Relativity},} \apj, 922, 149, \dodoi{10.3847/1538-4357/ac222a}

\bibitem[{D. {Deb} {et~al.}(2022){Deb}, {Mukhopadhyay}, \& {Weber}}]{deb}
{Deb}, D., {Mukhopadhyay}, B., \& {Weber}, F. 2022,
  \bibinfo{title}{{Anisotropic Magnetized White Dwarfs: Unifying Under- and
  Overluminous Peculiar and Standard Type Ia Supernovae},} \apj, 926, 66,
  \dodoi{10.3847/1538-4357/ac410b}

\bibitem[{P.~P. {Eggleton}(1971){Eggleton}}]{eggleton}
{Eggleton}, P.~P. 1971, \bibinfo{title}{{The evolution of low mass stars},}
  \mnras, 151, 351, \dodoi{10.1093/mnras/151.3.351}

\bibitem[{P.~P. {Eggleton}(1972){Eggleton}}]{eggleton2}
{Eggleton}, P.~P. 1972, \bibinfo{title}{{Composition changes during stellar
  evolution},} \mnras, 156, 361, \dodoi{10.1093/mnras/156.3.361}

\bibitem[{L. {Ferrario} {et~al.}(2020){Ferrario}, {Wickramasinghe}, \&
  {Kawka}}]{ferrario}
{Ferrario}, L., {Wickramasinghe}, D., \& {Kawka}, A. 2020,
  \bibinfo{title}{{Magnetic fields in isolated and interacting white dwarfs},}
  Advances in Space Research, 66, 1025, \dodoi{10.1016/j.asr.2019.11.012}

\bibitem[{R.~J. {Foley} {et~al.}(2013){Foley}, {Challis}, {Chornock},
  {Ganeshalingam}, {Li}, {Marion}, {Morrell}, {Pignata}, {Stritzinger},
  {Silverman}, {Wang}, {Anderson}, {Filippenko}, {Freedman}, {Hamuy}, {Jha},
  {Kirshner}, {McCully}, {Persson}, {Phillips}, {Reichart}, \&
  {Soderberg}}]{Iax}
{Foley}, R.~J., {Challis}, P.~J., {Chornock}, R., {et~al.} 2013,
  \bibinfo{title}{{Type Iax Supernovae: A New Class of Stellar Explosion},}
  \apj, 767, 57, \dodoi{10.1088/0004-637X/767/1/57}

\bibitem[{B. Franzon \& S. Schramm(2015)Franzon \& Schramm}]{franzon_rotwd}
Franzon, B., \& Schramm, S. 2015, \bibinfo{title}{{Effects of strong magnetic
  fields and rotation on white dwarf structure},} Phys. Rev. D, 92, 083006,
  \dodoi{10.1103/PhysRevD.92.083006}

\bibitem[{P. {Goldreich} \& A. {Reisenegger}(1992){Goldreich} \&
  {Reisenegger}}]{goldreich}
{Goldreich}, P., \& {Reisenegger}, A. 1992, \bibinfo{title}{{Magnetic Field
  Decay in Isolated Neutron Stars},} \apj, 395, 250, \dodoi{10.1086/171646}

\bibitem[{A. {Gupta} {et~al.}(2020){Gupta}, {Mukhopadhyay}, \& {Tout}}]{abhay}
{Gupta}, A., {Mukhopadhyay}, B., \& {Tout}, C.~A. 2020,
  \bibinfo{title}{{Suppression of luminosity and mass-radius relation of highly
  magnetized white dwarfs},} \mnras, 496, 894, \dodoi{10.1093/mnras/staa1575}

\bibitem[{I. {Hachisu} {et~al.}(2012){Hachisu}, {Kato}, \& {Nomoto}}]{hachisu}
{Hachisu}, I., {Kato}, M., \& {Nomoto}, K. 2012, \bibinfo{title}{{Final Fates
  of Rotating White Dwarfs and Their Companions in the Single Degenerate Model
  of Type Ia Supernovae},} \apjl, 756, L4, \dodoi{10.1088/2041-8205/756/1/L4}

\bibitem[{C.~R. Harris {et~al.}(2020)Harris, Millman, van~der Walt, Gommers,
  Virtanen, Cournapeau, Wieser, Taylor, Berg, Smith, Kern, Picus, Hoyer, van
  Kerkwijk, Brett, Haldane, del R{'{\i}}o, Wiebe, Peterson,
  G{'{e}}rard-Marchant, Sheppard, Reddy, Weckesser, Abbasi, Gohlke, \&
  Oliphant}]{numpy}
Harris, C.~R., Millman, K.~J., van~der Walt, S.~J., {et~al.} 2020,
  \bibinfo{title}{Array programming with {NumPy},} Nature, 585, 357,
  \dodoi{10.1038/s41586-020-2649-2}

\bibitem[{J.~S. {Heyl} \& S.~R. {Kulkarni}(1998){Heyl} \&
  {Kulkarni}}]{heylkulkarni}
{Heyl}, J.~S., \& {Kulkarni}, S.~R. 1998, \bibinfo{title}{{How Common Are
  Magnetars? The Consequences of Magnetic Field Decay},} \apjl, 506, L61,
  \dodoi{10.1086/311628}

\bibitem[{Y. {Hillman} {et~al.}(2016){Hillman}, {Prialnik}, {Kovetz}, \&
  {Shara}}]{flash}
{Hillman}, Y., {Prialnik}, D., {Kovetz}, A., \& {Shara}, M.~M. 2016,
  \bibinfo{title}{{Growing White Dwarfs to the Chandrasekhar Limit: The
  Parameter Space of the Single Degenerate SNIa Channel},} \apj, 819, 168,
  \dodoi{10.3847/0004-637X/819/2/168}

\bibitem[{S. {H{\"o}fner} \& H. {Olofsson}(2018){H{\"o}fner} \&
  {Olofsson}}]{agb_rate}
{H{\"o}fner}, S., \& {Olofsson}, H. 2018, \bibinfo{title}{{Mass loss of stars
  on the asymptotic giant branch. Mechanisms, models and measurements},} \aapr,
  26, 1, \dodoi{10.1007/s00159-017-0106-5}

\bibitem[{D.~A. {Howell} {et~al.}(2006){Howell}, {Sullivan}, {Nugent}, {Ellis},
  {Conley}, {Le Borgne}, {Carlberg}, {Guy}, {Balam}, {Basa}, {Fouchez}, {Hook},
  {Hsiao}, {Neill}, {Pain}, {Perrett}, \& {Pritchet}}]{SN2003fg}
{Howell}, D.~A., {Sullivan}, M., {Nugent}, P.~E., {et~al.} 2006,
  \bibinfo{title}{{The type Ia supernova SNLS-03D3bb from a
  super-Chandrasekhar-mass white dwarf star},} \nat, 443, 308,
  \dodoi{10.1038/nature05103}

\bibitem[{J.~D. Hunter(2007)Hunter}]{matplotlib}
Hunter, J.~D. 2007, \bibinfo{title}{Matplotlib: A 2D graphics environment,}
  Computing In Science \& Engineering, 9, 90

\bibitem[{N. {Hurley-Walker} {et~al.}(2023){Hurley-Walker}, {Rea}, {McSweeney},
  {Meyers}, {Lenc}, {Heywood}, {Hyman}, {Men}, {Clarke}, {Coti Zelati},
  {Price}, {Horv{\'a}th}, {Galvin}, {Anderson}, {Bahramian}, {Barr}, {Bhat},
  {Caleb}, {Dall'Ora}, {de Martino}, {Giacintucci}, {Morgan}, {Rajwade},
  {Stappers}, \& {Williams}}]{gpm}
{Hurley-Walker}, N., {Rea}, N., {McSweeney}, S.~J., {et~al.} 2023,
  \bibinfo{title}{{A long-period radio transient active for three decades},}
  \nat, 619, 487, \dodoi{10.1038/s41586-023-06202-5}

\bibitem[{O. {Ironi} {et~al.}(2025){Ironi}, {Ben-Ami}, {Hallakoun}, \&
  {Shahaf}}]{IfmR_Binaries}
{Ironi}, O., {Ben-Ami}, S., {Hallakoun}, N., \& {Shahaf}, S. 2025,
  \bibinfo{title}{{The Initial-to-final Mass Relation of White Dwarfs in
  Intermediate-separation Binaries},} \apj, 982, 20,
  \dodoi{10.3847/1538-4357/adb5f2}

\bibitem[{S. Kalita \& B. Mukhopadhyay(2018)Kalita \&
  Mukhopadhyay}]{surajit2018}
Kalita, S., \& Mukhopadhyay, B. 2018, \bibinfo{title}{{Modified Einstein's
  gravity to probe the sub- and super-Chandrasekhar limiting mass white dwarfs:
  a new perspective to unify under- and over-luminous type Ia supernovae},}
  JCAP, 09, 007, \dodoi{10.1088/1475-7516/2018/09/007}

\bibitem[{S. Kalita \& B. Mukhopadhyay(2019)Kalita \&
  Mukhopadhyay}]{surajit2019}
Kalita, S., \& Mukhopadhyay, B. 2019, \bibinfo{title}{{Continuous gravitational
  wave from magnetized white dwarfs and neutron stars: possible missions for
  LISA, DECIGO, BBO, ET detectors},} Mon. Not. Roy. Astron. Soc., 490, 2692,
  \dodoi{10.1093/mnras/stz2734}

\bibitem[{D. {Karinkuzhi} {et~al.}(2024){Karinkuzhi}, {Mukhopadhyay},
  {Wickramasinghe}, \& {Tout}}]{drisya}
{Karinkuzhi}, D., {Mukhopadhyay}, B., {Wickramasinghe}, D., \& {Tout}, C.~A.
  2024, \bibinfo{title}{{Mass-radius relation for magnetized white dwarfs from
  SDSS},} \mnras, 529, 4577, \dodoi{10.1093/mnras/stae829}

\bibitem[{H. {Komatsu} {et~al.}(1989){Komatsu}, {Eriguchi}, \&
  {Hachisu}}]{komatsu}
{Komatsu}, H., {Eriguchi}, Y., \& {Hachisu}, I. 1989, \bibinfo{title}{{Rapidly
  rotating general relativistic stars. I - Numerical method and its application
  to uniformly rotating polytropes},} \mnras, 237, 355,
  \dodoi{10.1093/mnras/237.2.355}

\bibitem[{H. Liu {et~al.}(2014)Liu, Zhang, \& Wen}]{charged}
Liu, H., Zhang, X., \& Wen, D. 2014, \bibinfo{title}{{One possible solution of
  peculiar type Ia supernovae explosions caused by a charged white dwarf},}
  Phys. Rev. D, 89, 104043, \dodoi{10.1103/PhysRevD.89.104043}

\bibitem[{R.~G. {Martin} {et~al.}(2006){Martin}, {Tout}, \&
  {Lesaffre}}]{Cignition}
{Martin}, R.~G., {Tout}, C.~A., \& {Lesaffre}, P. 2006,
  \bibinfo{title}{{Critical mass transfer in double-degenerate Type Ia
  supernovae},} \mnras, 373, 263, \dodoi{10.1111/j.1365-2966.2006.11019.x}

\bibitem[{A. {Mastrano} {et~al.}(2011){Mastrano}, {Melatos}, {Reisenegger}, \&
  {Akg{\"u}n}}]{ellipticity}
{Mastrano}, A., {Melatos}, A., {Reisenegger}, A., \& {Akg{\"u}n}, T. 2011,
  \bibinfo{title}{{Gravitational wave emission from a magnetically deformed
  non-barotropic neutron star},} \mnras, 417, 2288,
  \dodoi{10.1111/j.1365-2966.2011.19410.x}

\bibitem[{B. {Mukhopadhyay} \& A.~R. {Rao}(2016){Mukhopadhyay} \&
  {Rao}}]{bm_sgraxp}
{Mukhopadhyay}, B., \& {Rao}, A.~R. 2016, \bibinfo{title}{{Soft gamma-ray
  repeaters and anomalous X-ray pulsars as highly magnetized white dwarfs},}
  \jcap, 2016, 007, \dodoi{10.1088/1475-7516/2016/05/007}

\bibitem[{B. {Mukhopadhyay} {et~al.}(2017){Mukhopadhyay}, {Rao}, \&
  {Bhatia}}]{arsco_bm}
{Mukhopadhyay}, B., {Rao}, A.~R., \& {Bhatia}, T.~S. 2017, \bibinfo{title}{{AR
  Sco as a possible seed of highly magnetized white dwarf},} \mnras, 472, 3564,
  \dodoi{10.1093/mnras/stx2119}

\bibitem[{B. {Mukhopadhyay} {et~al.}(2021){Mukhopadhyay}, {Sarkar}, \&
  {Tout}}]{BMVirial}
{Mukhopadhyay}, B., {Sarkar}, A., \& {Tout}, C.~A. 2021,
  \bibinfo{title}{{Modified virial theorem for highly magnetized white
  dwarfs},} \mnras, 500, 763, \dodoi{10.1093/mnras/staa3136}

\bibitem[{P. {Neunteufel} {et~al.}(2019){Neunteufel}, {Yoon}, \&
  {Langer}}]{he_acc1}
{Neunteufel}, P., {Yoon}, S.~C., \& {Langer}, N. 2019,
  \bibinfo{title}{{Evolution of helium star plus carbon-oxygen white dwarf
  binary systems and implications for diverse stellar transients and
  hypervelocity stars},} \aap, 627, A14, \dodoi{10.1051/0004-6361/201935322}

\bibitem[{R. Nityananda \& S. Konar(2014)Nityananda \& Konar}]{nitya}
Nityananda, R., \& Konar, S. 2014, \bibinfo{title}{{Strong constraints on
  magnetized white dwarfs surpassing the Chandrasekhar mass limit},} Phys. Rev.
  D, 89, 103017, \dodoi{10.1103/PhysRevD.89.103017}

\bibitem[{J.~P. {Ostriker} \& F.~D.~A. {Hartwick}(1968){Ostriker} \&
  {Hartwick}}]{ostriker}
{Ostriker}, J.~P., \& {Hartwick}, F.~D.~A. 1968, \bibinfo{title}{{Rapidly
  Rotating Stars.IV. Magnetic White Dwarfs},} \apj, 153, 797,
  \dodoi{10.1086/149706}

\bibitem[{E. {Otoniel} {et~al.}(2025){Otoniel}, {Pretel}, {Mota}, {Flores}, \&
  {Alves}}]{fRT_gravity}
{Otoniel}, E., {Pretel}, J. M.~Z., {Mota}, C.~E., {Flores}, C. O.~V., \&
  {Alves}, V. B.~T. 2025, \bibinfo{title}{{White dwarf structure in
  $f(R,T,L_m)$ gravity: beyond the Chandrasekhar mass limit},} arXiv e-prints,
  arXiv:2507.18745, \dodoi{10.48550/arXiv.2507.18745}

\bibitem[{R. {Pakmor} {et~al.}(2021){Pakmor}, {Zenati}, {Perets}, \&
  {Toonen}}]{HeCO_WD}
{Pakmor}, R., {Zenati}, Y., {Perets}, H.~B., \& {Toonen}, S. 2021,
  \bibinfo{title}{{Thermonuclear explosion of a massive hybrid HeCO white dwarf
  triggered by a He detonation on a companion},} \mnras, 503, 4734,
  \dodoi{10.1093/mnras/stab686}

\bibitem[{S.~G. {Parsons} {et~al.}(2017){Parsons}, {G{\"a}nsicke}, {Marsh},
  {Ashley}, {Bours}, {Breedt}, {Burleigh}, {Copperwheat}, {Dhillon}, {Green},
  {Hardy}, {Hermes}, {Irawati}, {Kerry}, {Littlefair}, {McAllister},
  {Rattanasoon}, {Rebassa-Mansergas}, {Sahman}, \& {Schreiber}}]{parsons}
{Parsons}, S.~G., {G{\"a}nsicke}, B.~T., {Marsh}, T.~R., {et~al.} 2017,
  \bibinfo{title}{{Testing the white dwarf mass-radius relationship with
  eclipsing binaries},} \mnras, 470, 4473, \dodoi{10.1093/mnras/stx1522}

\bibitem[{S. {Perlmutter} {et~al.}(1999){Perlmutter}, {Aldering}, {Goldhaber},
  {Knop}, {Nugent}, {Castro}, {Deustua}, {Fabbro}, {Goobar}, {Groom}, {Hook},
  {Kim}, {Kim}, {Lee}, {Nunes}, {Pain}, {Pennypacker}, {Quimby}, {Lidman},
  {Ellis}, {Irwin}, {McMahon}, {Ruiz-Lapuente}, {Walton}, {Schaefer}, {Boyle},
  {Filippenko}, {Matheson}, {Fruchter}, {Panagia}, {Newberg}, {Couch}, \&
  {Project}}]{darkenergy2}
{Perlmutter}, S., {Aldering}, G., {Goldhaber}, G., {et~al.} 1999,
  \bibinfo{title}{{Measurements of {\ensuremath{\Omega}} and
  {\ensuremath{\Lambda}} from 42 High-Redshift Supernovae},} \apj, 517, 565,
  \dodoi{10.1086/307221}

\bibitem[{M. {Perpiny{\`a}-Vall{\`e}s} {et~al.}(2019){Perpiny{\`a}-Vall{\`e}s},
  {Rebassa-Mansergas}, {G{\"a}nsicke}, {Toonen}, {Hermes}, {Gentile Fusillo},
  \& {Tremblay}}]{tripleWD}
{Perpiny{\`a}-Vall{\`e}s}, M., {Rebassa-Mansergas}, A., {G{\"a}nsicke}, B.~T.,
  {et~al.} 2019, \bibinfo{title}{{Discovery of the first resolved triple white
  dwarf},} \mnras, 483, 901, \dodoi{10.1093/mnras/sty3149}

\bibitem[{M.~M. {Phillips}(1993){Phillips}}]{philips}
{Phillips}, M.~M. 1993, \bibinfo{title}{{The Absolute Magnitudes of Type IA
  Supernovae},} \apjl, 413, L105, \dodoi{10.1086/186970}

\bibitem[{L. {Piersanti} {et~al.}(2014){Piersanti}, {Tornamb{\'e}}, \&
  {Yungelson}}]{heflash}
{Piersanti}, L., {Tornamb{\'e}}, A., \& {Yungelson}, L.~R. 2014,
  \bibinfo{title}{{He-accreting white dwarfs: accretion regimes and final
  outcomes},} \mnras, 445, 3239, \dodoi{10.1093/mnras/stu1885}

\bibitem[{O.~R. {Pols} {et~al.}(1995){Pols}, {Tout}, {Eggleton}, \&
  {Han}}]{pols}
{Pols}, O.~R., {Tout}, C.~A., {Eggleton}, P.~P., \& {Han}, Z. 1995,
  \bibinfo{title}{{Approximate input physics for stellar modelling},} \mnras,
  274, 964, \dodoi{10.1093/mnras/274.3.964}

\bibitem[{Y. Qu \& B. Zhang(2025)Qu \& Zhang}]{wdp3}
Qu, Y., \& Zhang, B. 2025, \bibinfo{title}{{Magnetic Interactions in White
  Dwarf Binaries as Mechanism for Long-period Radio Transients},} Astrophys.
  J., 981, 34, \dodoi{10.3847/1538-4357/adb1b5}

\bibitem[{L.~G. {Quentin} \& C.~A. {Tout}(2018){Quentin} \&
  {Tout}}]{quentin2018}
{Quentin}, L.~G., \& {Tout}, C.~A. 2018, \bibinfo{title}{{Rotation and
  magnetism in intermediate-mass stars},} \mnras, 477, 2298,
  \dodoi{10.1093/mnras/sty770}

\bibitem[{A. {Ravi} {et~al.}(2025){Ravi}, {Govindarajan}, \&
  {Kalita}}]{trg_hubble}
{Ravi}, A., {Govindarajan}, T.~R., \& {Kalita}, S. 2025,
  \bibinfo{title}{{Over-Luminous Type Ia Supernovae and Standard Candle
  Cosmology},} arXiv e-prints, arXiv:2503.13904,
  \dodoi{10.48550/arXiv.2503.13904}

\bibitem[{N. Rea {et~al.}(2024)Rea {et~al.}}]{wdp2}
Rea, N., {et~al.} 2024, \bibinfo{title}{{Long-period Radio Pulsars: Population
  Study in the Neutron Star and White Dwarf Rotating Dipole Scenarios},}
  Astrophys. J., 961, 214, \dodoi{10.3847/1538-4357/ad165d}

\bibitem[{A.~G. {Riess} {et~al.}(1998){Riess}, {Filippenko}, {Challis},
  {Clocchiatti}, {Diercks}, {Garnavich}, {Gilliland}, {Hogan}, {Jha},
  {Kirshner}, {Leibundgut}, {Phillips}, {Reiss}, {Schmidt}, {Schommer},
  {Smith}, {Spyromilio}, {Stubbs}, {Suntzeff}, \& {Tonry}}]{darkenergy1}
{Riess}, A.~G., {Filippenko}, A.~V., {Challis}, P., {et~al.} 1998,
  \bibinfo{title}{{Observational Evidence from Supernovae for an Accelerating
  Universe and a Cosmological Constant},} \aj, 116, 1009,
  \dodoi{10.1086/300499}

\bibitem[{H. {Saio} \& K. {Nomoto}(1985){Saio} \& {Nomoto}}]{nomoto_ns}
{Saio}, H., \& {Nomoto}, K. 1985, \bibinfo{title}{{Evolution of a merging pair
  of C + O white dwarfs to form a single neutron star},} \aap, 150, L21

\bibitem[{H. Saio \& K. Nomoto(2004)Saio \& Nomoto}]{saio_offcenterC}
Saio, H., \& Nomoto, K. 2004, \bibinfo{title}{{Off - center carbon ignition in
  rapidly rotating, accreting carbon - oxygen white dwarfs},} Astrophys. J.,
  615, 444, \dodoi{10.1086/423976}

\bibitem[{N. Soker \& E. Bear(2023)Soker \& Bear}]{core_degen}
Soker, N., \& Bear, E. 2023, \bibinfo{title}{{The core degenerate scenario for
  the type Ia supernova SN 2020eyj},} Mon. Not. Roy. Astron. Soc., 521, 4561,
  \dodoi{10.1093/mnras/stad798}

\bibitem[{S. Subramanian \& B. Mukhopadhyay(2015)Subramanian \&
  Mukhopadhyay}]{sathya}
Subramanian, S., \& Mukhopadhyay, B. 2015, \bibinfo{title}{{GRMHD formulation
  of highly super-Chandrasekhar rotating magnetized white dwarfs: stable
  configurations of non-spherical white dwarfs},} Mon. Not. Roy. Astron. Soc.,
  454, 752, \dodoi{10.1093/mnras/stv1983}

\bibitem[{S. {Taubenberger}(2017){Taubenberger}}]{Taubenberger2017}
{Taubenberger}, S. 2017, \bibinfo{title}{{The Extremes of Thermonuclear
  Supernovae},} in Handbook of Supernovae, ed. A.~W. {Alsabti} \& P.~{Murdin},
  317, \dodoi{10.1007/978-3-319-21846-5_37}

\bibitem[{S. Taubenberger {et~al.}(2008)Taubenberger {et~al.}}]{underlum}
Taubenberger, S., {et~al.} 2008, \bibinfo{title}{{The underluminous Type Ia
  Supernova 2005bl and the class of objects similar to SN 1991bg},} Mon. Not.
  Roy. Astron. Soc., 385, 75, \dodoi{10.1111/j.1365-2966.2008.12843.x}

\bibitem[{S. {Taubenberger} {et~al.}(2011){Taubenberger}, {Benetti},
  {Childress}, {Pakmor}, {Hachinger}, {Mazzali}, {Stanishev}, {Elias-Rosa},
  {Agnoletto}, {Bufano}, {Ergon}, {Harutyunyan}, {Inserra}, {Kankare},
  {Kromer}, {Navasardyan}, {Nicolas}, {Pastorello}, {Prosperi}, {Salgado},
  {Sollerman}, {Stritzinger}, {Turatto}, {Valenti}, \&
  {Hillebrandt}}]{sn2009dc}
{Taubenberger}, S., {Benetti}, S., {Childress}, M., {et~al.} 2011,
  \bibinfo{title}{{High luminosity, slow ejecta and persistent carbon lines: SN
  2009dc challenges thermonuclear explosion scenarios},} \mnras, 412, 2735,
  \dodoi{10.1111/j.1365-2966.2010.18107.x}

\bibitem[{F.~X. {Timmes} {et~al.}(1994){Timmes}, {Woosley}, \& {Taam}}]{timmes}
{Timmes}, F.~X., {Woosley}, S.~E., \& {Taam}, R.~E. 1994, \bibinfo{title}{{The
  Conductive Propagation of Nuclear Flames. II. Convectively Bounded Flames in
  C+O and O+Ne+Mg Cores},} \apj, 420, 348, \dodoi{10.1086/173565}

\bibitem[{B. {Wang}(2018){Wang}}]{bowang_nova}
{Wang}, B. 2018, \bibinfo{title}{{Mass-accreting white dwarfs and type Ia
  supernovae},} Research in Astronomy and Astrophysics, 18, 049,
  \dodoi{10.1088/1674-4527/18/5/49}

\bibitem[{D.~T. {Wickramasinghe} {et~al.}(2014){Wickramasinghe}, {Tout}, \&
  {Ferrario}}]{tout_mostmagnstars}
{Wickramasinghe}, D.~T., {Tout}, C.~A., \& {Ferrario}, L. 2014,
  \bibinfo{title}{{The most magnetic stars},} \mnras, 437, 675,
  \dodoi{10.1093/mnras/stt1910}

\bibitem[{C. {Wu} {et~al.}(2019){Wu}, {Wang}, \& {Liu}}]{dd_sn1a}
{Wu}, C., {Wang}, B., \& {Liu}, D. 2019, \bibinfo{title}{{The outcomes of
  carbon-oxygen white dwarfs accreting CO-rich material},} \mnras, 483, 263,
  \dodoi{10.1093/mnras/sty3176}

\bibitem[{B. Zhang \& J. Gil(2005)Zhang \& Gil}]{wdp1}
Zhang, B., \& Gil, J. 2005, \bibinfo{title}{{GCRT J1745-3009 as a transient
  white dwarf pulsar},} Astrophys. J. Lett., 631, L143, \dodoi{10.1086/497428}

\bibitem[{Z. Zuraiq {et~al.}(2026)Zuraiq, Das, Deb, Kalita, Weber, \&
  Mukhopadhyay}]{UniversePaper}
Zuraiq, Z., Das, M., Deb, D., {et~al.} 2026, \bibinfo{title}{{Anisotropic
  Compact Stars: Theory and Simulation from Microphysical Models to Macroscopic
  Structure and Observables},} Universe, 12, 130,
  \dodoi{10.3390/universe12050130}

\bibitem[{Z. Zuraiq \& B. Mukhopadhyay(2026)Zuraiq \& Mukhopadhyay}]{zz_csc}
Zuraiq, Z., \& Mukhopadhyay, B. 2026, \bibinfo{title}{{Anisotropic hybrid
  stars: Interplay of superconductivity and magnetic field leading to
  gravitational waves},} \doarXiv{2604.06308}

\bibitem[{Z. {Zuraiq} {et~al.}(2024){Zuraiq}, {Mukhopadhyay}, \& {Weber}}]{zz}
{Zuraiq}, Z., {Mukhopadhyay}, B., \& {Weber}, F. 2024, \bibinfo{title}{{Massive
  neutron stars as mass gap candidates: Exploring equation of state and
  magnetic field},} \prd, 109, 023027, \dodoi{10.1103/PhysRevD.109.023027}

\end{thebibliography}
\bibliographystyle{aasjournalv7}



\end{document}